\documentstyle[11pt,epsf]{article}

\input psfig
\def\beq{\begin{eqnarray}}
\def\eeq{\end{eqnarray}}
\def\bea{\begin{eqnarray*}}
\def\eea{\end{eqnarray*}}

\def\centeron#1#2{{\setbox0=\hbox{#1}\setbox1=\hbox{#2}\ifdim
\wd1>\wd0\kern.5\wd1\kern-.5\wd0\fi
\copy0\kern-.5\wd0\kern-.5\wd1\copy1\ifdim\wd0>\wd1
\kern.5\wd0\kern-.5\wd1\fi}}
\def\ltap{\;\centeron{\raise.35ex\hbox{$<$}}{\lower.65ex\hbox{$\sim$}}\;}
\def\gtap{\;\centeron{\raise.35ex\hbox{$>$}}{\lower.65ex\hbox{$\sim$}}\;}

\def\singleandthirdspaced{\baselineskip=\normalbaselineskip\multiply
    \baselineskip by 130\divide\baselineskip by 100}
\def\singlespaced{\baselineskip=\normalbaselineskip}

\newcommand{\newc}{\newcommand}
\newc{\qbar}{{\overline q}}
\newc{\Kahler}{K\"ahler }
\newc{\deltaGS}{\delta_{\rm GS}}
\begin{document}
\begin{titlepage}
\begin{flushright}
{\large
hep-th/9805007 \\
SCIPP-98/15\\

}
\end{flushright}

\vskip 1.2cm

\begin{center}

{\LARGE\bf Renormalization of Higher Derivative Operators
in the Matrix Model}

\vskip 1.4cm

{\large
Michael Dine, Robert Echols and Joshua P. Gray}
\\
\vskip 0.4cm
{\it Santa Cruz Institute for Particle Physics,
     Santa Cruz CA 95064  } \\

\vskip 4pt

\vskip 1.5cm

\begin{abstract} 
$M$-theory is believed to be described in various dimensions by
large $N$ field theories.  It has been further conjectured that
at finite $N$, these theories describe the discrete light cone
quantization (DLCQ) of $M$ theory.  Even at low energies, this is not
necessarily the same thing as the DLCQ of supergravity.  It is
believed that this is only the case for quantities which are
protected by non-renormalization theorems.  In $0+1$ and $1+1$
dimensions, we provide further evidence of a non-renormalization
theorem for the $v^4$ terms, but also give evidence that there are not
such theorems at order $v^8$ and higher.  These results are
compatible with known facts about the matrix model.
\end{abstract}

\end{center}

\vskip 1.0 cm

\end{titlepage}
\setcounter{footnote}{0}
\setcounter{page}{2}
\setcounter{section}{0}
\setcounter{subsection}{0}
\setcounter{subsubsection}{0}

\singleandthirdspaced

\section{Introduction}

The matrix model conjecture has two parts.  In its original
form, it was suggested that in the large $N$ limit, the matrix
model should describe $M$ theory in the infinite momentum
frame\cite{bfss,reviews}.
Susskind proposed that for finite $N$, the matrix model
describes the discrete light cone quantization of $M$
theory\cite{dlcq}.
Naively, one might take this to mean that for large distance
processes, the predictions of the matrix model should agree with
the DLCQ
of $11$-dimensional supergravity.
Most of the initial matrix model tests seemed to support this
hypothesis.  For example, graviton-graviton scattering agrees
at finite $N$.  We will refer to this expectation as the
``naive DLCQ.''

Seiberg and Sen have subsequently given a derivation of the
finite $N$ version of the matrix model\cite{seibergdlcq,sendlcq}.
Their argument
begins with the observation that the DLCQ, in general,
can be thought of as a large boost of a theory compactified
on a small, space-like circle.  However,
their argument does not necessarily establish that amplitudes
in the matrix model must agree with the
naive DLCQ.  As pointed out by Polchinski
and Hellerman, this derivation implies that there is a small space-like
distance in the problem, and so the DLCQ does not describe
some simple, low energy limit of string
theory\cite{ph}.
Thorn\cite{thorn}
has stressed that even ignoring the zero modes, in field
theory the DLCQ only ``works" for generic
momenta; for special kinematics (e.g. zero longitudinal
momentum exchange, the kinematical
situation most easily studied in the matrix model) there
are subtle issues.  Indeed, the matrix model
does not reproduce the naive DLCQ for
for scattering in non-trivial
backgrounds\cite{ooguri}.

The situation has been well-summarized in \cite{banksreview}.
It raises the question:  why does the finite $N$ model
agree with the naive DLCQ sometimes?   Consider the case of graviton-graviton
scattering.  In the matrix model, the relevant
terms are the $v^4$ terms in the effective
action, i.e. the terms with four derivatives.
It is believed that there is a non-renormalization
theorem for these, so the lowest order term
must reproduce the exact answer,
explaining the agreement.  The evidence for this theorem
is limited.  In four dimensions, it has been proven
at infinite volume\cite{dsnr}.
In
$0+1$ dimensions, it has been shown by explicit
computation that there is no renormalization of the
$v^4$ terms at two loops in the case of $SU(2)$\cite{beckers}.

If $n$ graviton scattering is to work similarly, there
should be a non-renormalization theorem
for the $v^{2n}$ terms in the action.  In \cite{bbpt},
certain terms
at order $v^6$, which are subleading in $N$,
were shown to agree with the DLCQ of
Einstein's gravity.  This suggests
that some quantities at order $v^{6}$ are not
renormalized.
On the other hand, it would be
suprising if there were an infinite series
of such non-renormalization theorems.
The work of \cite{ooguri} suggests that there might be a problem
at the level of four graviton scattering\cite{private}.  Investigations of this
question will appear elsewhere.

In these notes, we will first give some further evidence supporting
the non-renormalization theorem for $v^4$ in low dimensions,
and direct evidence that
there are not general non-renormalization theorems
for $v^8$ and higher powers of the velocity (we will not
be able to make a definite statement about $v^6$).
In doing so, the first question we have to ask
is:  ``non-renormalization
of what?''  In four dimensions we are used to
the idea that non-renormalization theorems are statements
about a Wilsonian effective action.  For example, the
non-renormalization theorem discussed
in \cite{dsnr} is derived by considering the $N=4$ theory
on its Coulomb branch, and studying the effective action obtained by
integrating out massive and high frequency modes.  In $0+1$
and $1+1$ dimensions (or in finite volume), however,
there is not a notion of a moduli
space in the same sense.  Instead, one must adopt a Born-Oppeheimer
treatment of the problem, thinking of holding the slow modes fixed
and solving for the dynamics of the fast modes.    

The approach of most authors has been to compute the
one particle irreducible
effective action, using conventional field theory rules.
Consider the case of $SU(2)$.
At a given
order in $v$, the loop expansion is formally an expansion in powers
of $1/r^3$ ($1/r^2$), in $0+1$ ($1+1$) dimensions\cite{bbpt}.  Here,
$r$ is
the expectation value of the adjoint fields (transverse
separation of the gravitons, in the matrix model interpretation).
The spectrum includes states with mass (frequency) of order $r$ and
massless states.
In the two-loop computation of \cite{beckers}, 
individual diagrams contributing to the
effective action containing massless states are infrared divergent.
The authors of this reference dealt with this by using dimensional
regularization, defining 
\beq
\int {d^d p \over p^2}=0.
\label{dimreg}
\eeq
With this regulator, these authors find that there is no
renormalization of the $v^4$ term.  The result involves not
only fermi-bose cancellations, but also cancellations between
diagrams containing only massive states and diagrams
containing massless states.

This result is encouraging, but
since infrared divergences usually signal real physics,
one might worry about the regularization procedure.  However, there
are many infrared divergent diagrams, and,
as we will see in section 2, in the case of $v^4$,
the infrared divergences cancel and there is
no sensitivity to the regularization procedure.  We will also see
that this cancellation is quite special to $v^4$, and
there is no reason to expect it to occur
for higher orders in velocity.

While it is true that we do not have a good definition of
a Wilsonian effective action,
for the success of the naive DLCQ, what really interests
us is the scattering amplitude.  For the success of the
naive DLCQ, at ${\cal O}(v^4)$,
we actually require that there should be no corrections
to this amplitude.  This is, as we will
explain in the next section, equivalent to the requirement
that there should be no corrections to the 1PI effective action. 
We will see that, for
two particle scattering, at order $v^8$ and higher,
there are infrared divergences in the perturbation theory
in powers of $1/r^3$.\footnote{In the
eikonal approximation appropriate to this
problem, terms in the effective action
of order $v^n$ translate into terms in the
scattering amplitude of order $v^{n-1}$.  We will
count powers of $v$ and $r$ in this paper
as appropriate for the effective action.}  As a result, we believe that
the expansion in powers of velocity breaks down.

\begin{figure}[htbp]
\centering
\centerline{\psfig{file=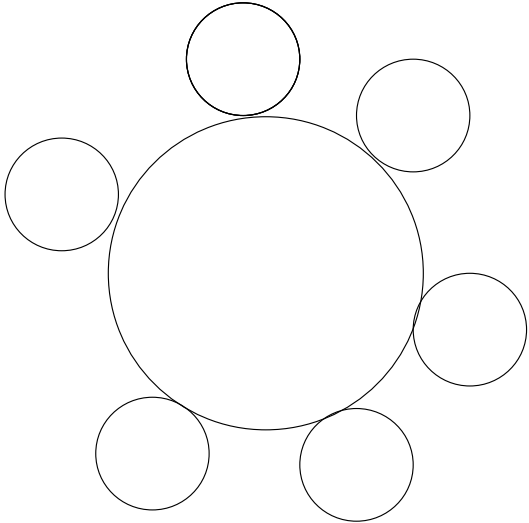,width=7cm}}
\caption{Infrared divergent contributions to the effective action.}
\label{oneloophiggs}
\end{figure}

The origin of the infrared problem is easily understood.
In $\ell$ loop
order, $\ell \ge 2$, consider the diagram
shown in fig. 1.  Here the
central loop contains a massive field, and the $\ell-1$
smaller loops contain massless fields.  In momentum
space, this graph is proportional to
\beq
{{1 \over r^{6+ 2(\ell )-d}}
(\int {d^d p \over p^2})^{\ell -1}.}
\label{lloopdiv}
\eeq
Alternatively, if the amplitudes are written in coordinate space,
the propagator is ambiguous; individual diagrams are proportional
to this ambiguity.
In the infrared
limit, one can think of the integral over the massive
states as generating a local operator, and the
massless integrals as giving the ``vacuum matrix
element'' of this operator.  This same
type of analysis can be performed for all of the
infrared divergent graphs.  For the $v^4$ terms, we will see
in the next section that this matrix element vanishes.
However, this cancellation depends crucially on the fact
that $1/r^7$ is the Green's function for the nine-dimensional
Laplace operator, and does not hold for higher powers of $v$.
$v^6$ turns out to be special as well, because the one loop contribution
vanishes\cite{bbpt}, and we
cannot make a definite statement.  However, starting at order $v^8$,
it is easy to see that there are divergences, they do not
cancel, and they become
more severe in each order of perturbation theory.

These divergences presumably signify that there is not an 
expansion of the effective action (in the matrix model) in
powers of $v$ and $1/r^3$.  The problem seems analogous
to difficulties encountered in finite temperature field theories,
where the divergence is cut off by some dynamical mass scale
(e.g. the Debye mass or the magnetic mass in the case of
finite temperature QCD).  In the present
case, there is an effective mass proportional
to $v^4/r^9$ ($v^4/r^8$) in $0+1$ ($1+1$) dimensions,
which can cut off the integrals.
If this is the relevant scale, the $0+1$ expansion breaks down
at order $v^6$; in $1+1$, the problem occurs at the level
of $v^8$.

In the case of $SU(3)$ (and higher
rank groups), one can make the problem somewhat
sharper by exhibiting certain
finite renormalizations.    In this case there are two (or more)
scales, $R$ and $r$.  As in \cite{arvind}, one can
consider a hierarchy of scales (impact
parameters,in the matrix model interpretation),
$R\gg r$.
Again, the diagrams
contributing to the effective action contain infrared divergent
terms.  But there are also finite terms which behave as
$(1/R^2 r)^{\ell-1}$.
It is easy to isolate these terms.
Diagrams such as those of fig. 1, where now the small loops
contain fields of mass $r$ and the big loop masses of order $R$,
are of the form
\beq
{{v^4 \over R^{6+ 2(\ell )-d}}
(\int
{d^d p \over p^2 + r^2})^{l-1} \sim {v^4 \over R^{6+2\ell-d}
r^{\ell-1-d}}}
\label{lloopsuthree}
\eeq
(in $1+1$, the $r$ dependence is logarithmic).
In the third section, we will see that there
is a cancellation of the most singular term at order $v^4$
for $\ell=2$.
Based on the results for $SU(2)$, it  seems quite plausible that this
cancellation persists to all orders.  From the perspective of the matrix
model, this is reassuring, since there would be no sensible
spacetime interpretation for such terms.  As for $SU(2)$, it is not
easy to decide what happens at order $v^6$,
but at order $v^8$ and beyond, it is a simple
matter to show that there are renormalizations.

However, to determine
the full implications of these results requires settling
some subtle issues.  In particular, for these low dimension
theories, the significance of the effective action
is not completely clear, obscured, as we have
noted, by infrared
and (related\footnote{We thank Nathan Seiberg for stressing
this connection to us.}) operator ordering questions.
We will offer some remarks on these issues, but will
not completely resolve them.  Whether there are discrepancies
between the matrix model and four graviton scattering,
as suggested by these remarks about $v^8$, will
be discussed elsewhere.



\section{Infrared Divergences in $SU(2)$}

Consider, first, the matrix model with $N=2$ in $d=0+1$ and
$1+1$.  We will
write the bosonic part in terms of a set of ``fields,'' $x^i$, $i=1,\dots 9$,
and a ``gauge boson,'' $A$.  All of these fields are $SU(2)$
matrices.  There are flat directions with $\vec x$ a diagonal
matrix,
\beq
{\vec x = \vec r \tau_3/2.}
\label{xvev}
\eeq
Correspondingly, there are a set
of massive modes (i.e. modes with frequencies proportional
to $r$) and massless modes.   At one loop, integrating out the
massive modes in
this model is well-known to generate an effective action,
whose leading bosonic term behaves as\cite{bacchas,bc} 
\beq
{{\cal L}_1= {15 \over 16}{v^4 \over r^7}.}
\label{vfourth}
\eeq

When considering the scattering amplitude, in a path
integral approach, one considers
\beq
\langle \vec x_f(t_f) \vert \vec x_i(t_i) \rangle
\eeq
where $\vec x_f$ and $\vec x_i$ are the eigenvalues
of $\vec x_3$, the diagonal component of the matrix.
Expanding $\vec x$ about the classical solution
\beq
\vec x(t) = \vec x_{cl} + \delta \vec x 
\eeq
$$~~~~~~
\vec  x_i + {\vec x_f-\vec x_i \over t_f -t_i}(t-t_i)
+ \delta \vec x(t)
$$
$$~~~~~~=\vec b + \vec v t + \delta \vec x$$
one studies the region of large $\vert \vec b \vert$, small
$\vert \vec v \vert$.  In this regime, the amplitude can
be expanded in powers of $\vec v$\cite{bacchas, matrixcalculations}.
At higher orders, as we have noted,
there is a serious potential for infrared
divergences.  In $1+1$ dimension, the problem is familiar
from string theory.  Written in a fourier decomposition, the
two dimensional massless propagator is:
\beq
{\langle x(\sigma)x(\sigma^{\prime}) \rangle =
\int d^2 k {e^{ik \cdot (\sigma-\sigma^{\prime})} \over k^2},}
\label{twodprop}
\eeq
which is ill defined.  Correspondingly, the coordinate
space expression is 
\beq
{\langle x(\sigma)x(\sigma^{\prime}) \rangle =
\ln(\sigma-\sigma^{\prime})^2 + {\rm constant}.}
\label{coordspace}
\eeq
In string theory, one only considers Green's functions of
translationally invariant combinations of operators,
and these are infrared finite; equivalently, they are
independent of the arbitrary constant.

In $0+1$ dimensions, the divergences are even more severe.
If we try to write a momentum (frequency) space propagator
we have
\beq
{\langle \delta x_i(t)\delta 
 x_j(t^{\prime}) \rangle =
\delta_{ij}\int d\omega {e^{-i\omega(t-t^{\prime})}\over \omega^2}}
\label{onedprop}
\eeq
which is linearly divergent.  Correspondingly, the coordinate
space Green's function is ambiguous (dropping the vector symbol):
\beq
{\langle \delta x(t) \delta x(0) \rangle =
a t\theta(t) -b t \theta(-t) + ct + d,}
\label{twodproptwo}
\eeq
with $a+b=1$.

When we say below that infrared divergences do
(or do not) cancel in $1+1$ or
$0+1$ dimension, we will mean that they cancel at the level
of momentum space expansions, or alternatively that the quantities
in question are
not sensitive to the ambiguities in the propagators.

\begin{figure}[htbp]
\centering
\centerline{\psfig{file=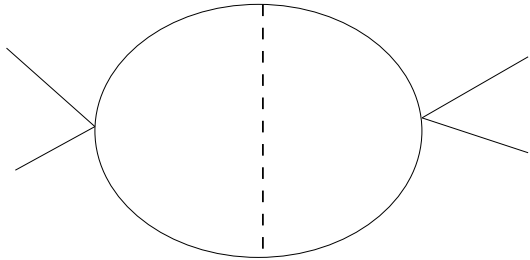,width=7cm,angle=-90}}
\caption{Some two loop corrections to the effective action.}
\label{oneloophiggstwo}
\end{figure}

Now consider two loop corrections to the effective
action.  Some sample diagrams
are shown in fig. 2.  Consider, in particular,
diagrams with one massive state and one massless state
running in the loop.  Individual diagrams with massless states in the loop
are infrared divergent, behaving as
\beq
{\int {d\omega \over \omega^2}}
\label{irdivergence}
\eeq
for small frequencies.   Note that the external
$x$'s must always attach to massive lines.
Because of this fact, and because the leading infrared divergence
always comes from such a small frequency region of integration,
the leading divergent piece of each diagram always factorizes into
a product of two one loop terms.  One is a massive loop,
with four external
``scalars'' ($x$'s), on which the time derivatives act, and two more without
derivatives.  The two
without derivatives are then
contracted with each other, forming
the massless loop.
In other words, the infrared divergent terms can all
be organized in terms of operators generated at one loop
of the form
\beq
{{\cal O}= v^4 \delta x^2/r^9.}
\label{oneloopops}
\eeq
The infrared divergence then arises from simply contracting
the two factors of $x$ in this expression, i.e. taking
the ``vacuum matrix element.''  

However, we do not need to compute all of the diagrams
to determine the coefficient of this term in the effective action!
In eqn. $5$, we can interpret $r^2$
as $(\vec x_{cl} + \delta \vec x)^2$, and expand in powers of $\delta \vec x$.
This gives
\beq
{\cal O}_1=
{{v^4 \over x_{cl}^7} \left (1 - {7 \over 2 x_{cl}^2}(2 \vec x_{cl} \cdot \vec
\delta x
+ \vec \delta x^2) +{7 \times 9 \over 8 x_{cl}^4}(2 \vec x_{cl} \cdot \vec
\delta x + \vec \delta x^2)^2 \right
).}
\label{sutwoexpansion}
\eeq
Taking the expectation value, the last
two (infrared divergent) terms in this expression cancel
because there are nine $x$'s.  A similar cancellation occurs in $1+1$
dimension.

It should be noted that there are no potential infrared divergences from
other diagrams.  Diagrams involving gauge fields (which exist in gauges
other than $A^o=0$) are not divergent.  The one loop effective action
must be gauge invariant, and this means that it must be independent of
the gauge field in $0+1$ dimensions, and involve at least two time derivatives
in $1+1$ dimension.  Diagrams involving fermions are not as divergent
due to the structure of the fermion propagator.

It is easy to extend this argument for the cancellation of the
most infrared singular terms
to higher orders.  At each order, the most singular
contribution comes from diagrams where several massless scalars
attach to a single loop of massive fields.  These diagrams correspond
to expanding the $1/r^7$ term to higher orders in $x$, and contracting
the $x$'s.  But $1/r^7$ is special, as it is the Green's function for the
nine-dimensional laplacian.  This means that, for $r \ne 0$, $r \gg x$,
\beq
{\nabla^2 {1 \over \vert \vec r + \vec \delta x \vert^7}=0} 
\label{laplacianzero}
\eeq
where the derivatives act with respect to $\vec r$.  Expanding in powers
of $\delta \vec x$, this must be true for every term in the sum.  It must also,
then, be true when we average over $x$.  But averaged over $\delta x$,
each term
is proportional to $1 \over r^n$ (times an infrared divergent integral).
So, except for the leading term, the coefficient of every other term in the
expansion must vanish, upon averaging.  The skeptical reader is
invited to check the next order explicitly.

Note that in the path integral framework, the
non-renormalization of the $v^4$ terms (and
the cancellation of ir divergences) in the effective
action immediately implies the same for the scattering amplitude.
Contributions which are not one particle irreducible are easily
shown to be of higher order in velocity.  The same holds true
for the $v^8$ terms which we discuss shortly.

Now consider higher orders in velocity.  At one loop,
there is no $v^6$ term in the effective action.  There is
a $v^8$ term,
\beq
{v^8 \over r^{15}}.
\label{veighth}
\eeq
Expanding the denominator as before, one now finds that there
is an infrared divergence at two loops.  Again, this cannot be
cancelled by graphs with fermions or gauge bosons.
While the expressions are ill defined, this presumably represents
a breakdown of the non-renormalization theorems.  The same statements
also hold in $1+1$ dimensions.

\begin{figure}[htbp]
\centering
\centerline{\psfig{file=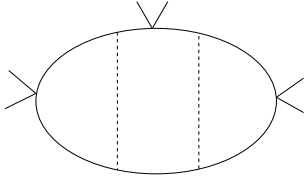,width=4cm,angle=-90}}
\caption{Three loop correction to the effective action.}
\label{oneloophiggsthree}
\end{figure}

Returning to the $v^6$ terms, as noted above, a $v^6$ term
is not generated at one loop.  Such a term is generated
at two loop\cite{bbpt}.  But we cannot simply apply our reasoning
to the two loop case.  The calculation of \cite{bbpt}
includes graphs with both massive and massless states.
At three loops, there are diagrams with zero, one or two
massless particles in the loop.  Expanding the two loop action
in powers of $\delta \vec x$, and contracting $<\vec \delta x \vec \delta x>$
correctly reproduces the infrared parts of diagrams with
one massless field, but double counts the diagrams with
two (see fig. 3).  So we cannot establish by this means
whether there is an infrared divergence (and a breakdown
of the non-renormalization theorem) for $SU(2)$ at
$v^6$.  This is just as well.  The fact that the calculation
of \cite{bbpt} successfully reproduces the naive DLCQ strongly
suggests that there is a non-renormalization theorem
for this case.     

Finally, we should note that the authors of \cite{beckers}\
have computed, using their regulator, the coefficient
of the $v^8$ term at two loops\cite{beckers2}.  However, they are not
able to perform a direct comparison with supergravity.

\section{Finite Renormalizations in $SU(3)$}

Consider, now, an $SU(3)$ gauge group.  In this case, taking
$x$ to be a $U(3)$ field, we will consider
``expectation values'' of the fields $x$ of the form:
\beq
{x^9 =\left ( \matrix{0 & 0 &  0 \cr 0 & r & 0
\cr 0 & 0 & R}
\right ).}
\label{suthreevev}
\eeq
This is not the most general expectation value, but it is sufficient for
our purposes.   In the language of $M$ theory or $D0$ branes,
this corresponds to three gravitons (branes) at locations $0, r$ and $R$
respectively.

Suppose that $r \ll R$.  Then there is an approximate
$SU(2)$ ($U(2)$) symmetry.
We can then imagine first integrating out
states with mass of order $R$, and then those with mass of order
$r$, to obtain an effective action for the massless fields.  At the first
step, we expect to generate an operator of the form
\beq
{{\cal O}_3=  {\vec v_3^4( a  \vec x^a \cdot
\vec x^a+ b  \hat R \cdot \vec x^a \hat R \cdot
\vec x^a)
 \over R^9},}
\label{suthreeoperator}
\eeq
where $\vec x_a$ are the $SU(2)$ triplet fields.
Then, replacing $x^a x^a$ by 
\beq
{ \langle x^a_i x^a_j \rangle = \delta_{ij}({1
 \over r} + \int {d \omega \over (2 \pi)\omega^2})       }
\label{twoptr}
\eeq
in this expression, we obtain a result proportional to
\beq
{{\cal O}_4 = {v_3^4 \over R^9 r},}
\label{mainoperator}
\eeq
as well as a potentially infrared divergent term.

As in the case of $SU(2)$, it is not difficult to verify that the coefficient
of ${\cal O}_4$, as well as the infrared divergence, vanishes to two loops.
In the $SU(3)$ case, the one-loop result is:          
\beq
{{\cal L}_{v^4} \propto
( {v_{12}^4 \over x_{12}^7}
+  {v_{13}^4 \over x_{13}^7}
+  {v_{23}^4 \over x_{23}^7})}.
\label{oneloopsuthree}
\eeq
Write
\beq
{\vec x_1 =  \vec x_1
~~~~~~~ \vec x_2 = \vec r +\vec x_2 ~~~~~~~~~\vec x_3 = \vec R + \vec x_3,}
\label{suthreefluctuations}
\eeq
and expand the last
two terms in powers of the fluctuations, $\vec x$,
keeping only the part proportional to $v_3^4$.
The first order terms are $SU(2)$ singlets (they are proportional
to $\vec x_1 + \vec x_2$).  The quadratic terms contain the
$SU(2)$ non-singlet fields, $\vec x_1-\vec x_2$.
These couplings can be generalized to the $SU(2)$ invariant
coupling,
\beq
{\delta {\cal L}={v^4 x^a x^a \over R^9}.}
\label{sutwoinvariant}
\eeq
Taking the expectation value, we see that
as in the case of the $SU(2)$ infrared divergences,
the leading $1/r$ and infrared divergent pieces cancel.  It is not so easy to
check higher orders, in this case, since one can't generalize,
e.g., the $\vec x^n$ terms unambiguously to $SU(2)$-invariant
expressions.  However, we have checked
explicitly the cancellation to next order,
and expect the same will occur for higher orders.

Equally important, repeating this analysis for $v_3^8$
and higher, we see that there will be $1/r$ corrections
at two loops for $v^8$.  Again, because of the vanishing
of the $v^6$ term at one loop, we cannot establish by
this sort of reasoning whether or not there are corrections
to the various $v^6$ operators at three loops.

One must ask if there are other operators which can
contribute to $1/r$ effects.  But it is again easy to see, on gauge invariance
grounds, that diagrams containing gauge bosons cannot contribute.
Diagrams involving fermions have the wrong
dependence on $R$ and $r$.

This argument can be extended to $1+1$ dimensions.
There is again no infrared problem at ${\cal O}(v^4)$, and no terms
which depend on $\ln(r/R)$ (the analog of
the $1/r$ terms in the $0+1$ dimensional case).  Infrared
divergences and finite renormalizations 
do again appear at ${\cal O}(v^8)$.


\section{Conclusions:  Implications for the Matrix Model}

At order $v^4$, we have
provided additional evidence that there are non-renormalization
theorems.   At order $v^6$, our method does permit one
to say whether or not there are renormalizations at three
loops.  This is consistent with the result of \cite{bbpt}\
that the naive DLCQ works for the $v^6$ terms
in two graviton scattering. For $v^8$ and beyond,
we have found that there are both finite
and infinite renormalizations.  This
suggests that in four graviton scattering and beyond,
there may be differences between the matrix model
and tree level supergravity predictions.  This is under
investigation, and will be reported elsewhere.

It is perhaps worth describing the three
graviton calculation of \cite{arvind}\
in the language we have used in this paper.  On the gravity
side, it was shown that a term in the scattering
amplitude behaves as $v_{12}^2v_{13}^2v_{23}^2
\over r^7 R^7$ where
$r$ and $R$ are the graviton separations
and $R \gg r$.  The claim of \cite{arvind}
is that there can be no terms on the
matrix model side of the form $v_3^4 v_{12}^2 \over
R^7 r^7$.  To see this,
one can consider the problem, as in the previous
section, by first integrating out states
of mass $R$.  At one loop, diagram by diagram, there are
terms proportional to $1/R$, $v^2/R^3$, and
$v^4/R^7$, where the $v$'s indicate
various tensor structures.  Because of the
non-renormalization theorems for the terms with zero
and two derivatives, 
the various contributions to the first two
sets of operators cancel.  Expanding the $v^4/R^7$, the
terms which can yield $v_3^4$ are of the form
$v_3^4 x^ax^a/R^9$ and $v_3^2 x_3 x_3 v^a v^a/R^9$,  as in eqn. 23.  
Now allowing some momenta (frequency) to flow in the various
external lines, and contracting the two
$x^a$ or $v^a$ factors, all terms with four
$v_3^4$ factors come with $1/R^9$.
It is worth stressing that this is not a diagram by
diagram statement.   Individual one loop
diagrams, for example, will generate operators
like $v_3^2 x_3^2 x^a x^a/R^7$.  Contracting the
$x^a$'s will then lead to terms containing
$v_3^4 v_{12}/R^7 r^7$.  But these terms
must cancel.  Indeed, in this paper, in considering
the infrared divergences, we have implicitly
assumed cancellations of this kind.  For example,
in $SU(2)$ at one loop, individual diagrams
contribute to operators like $1/r \sim {\delta x^2 \over r^3}$,
and these would give rise to additional infrared divergences.
But supersymmetry insures cancellation of such
operators.

The type of analysis we have performed here, which
relies heavily on the singular infrared behavior
of low dimensional field theories,
cannot be readily extended to higher dimensions.  In $2+1$
dimensions and beyond, rather than yielding an infrared divergence,
taking then vacuum matrix elements of operators like
${\cal O}_1$, ${\cal O}_3$
in the low energy theory yields ultraviolet divergences.


As we have stressed, the interpretation of the effective
action we have considered here is not completely clear,
and it is best to think in terms of calculating the
path integral for the scattering amplitude.  In the case of $SU(2)$,
we have seen that while the formal expansion of the path integral
amplitude in powers of velocity is potentially quite infrared
divergent, the most severe divergences
cancel through order $v^4$.  We have argued that
this will not persist beyond order $v^8$, and that
this signals a breakdown of the velocity expansion.
We have
also seen that in the case of $SU(3)$, large finite corrections
also cancel at order $v^4$, but this  is
not the case at ${\cal O}(v^8)$ and higher.

\noindent
{\bf Acknowledgements:}

\noindent
We thank W. Fischler, A. Rajaraman, L. Randall and N. Seiberg
for discussions.   This work supported in part by the U.S.
Department of Energy.

\noindent
{\bf Note added:}
Since this paper appeared on hep-th, there have been a number 
of developments.  First, it has been shown that
supersymmetry uniquely determines the $v^4$ term
in the quantum mechanics\cite{sethi}, which establishes
the non-renormalization theorem.  Still, one might worry
that in the presence of infrared divergences, there are
subtle issues in defining the effective action. The cancellation
we have exhibited provides some insight into this question.
A similar non-renormalization theorem has also
been proven for the $v^6$ terms in $SU(2)$\cite{sethitwo}.
Second, in the original version of this paper, it was assumed that
there were already discrepancies between the Matrix model
and naive DLCQ at the level of three graviton scattering.
The calculation of \cite{yoneya} shows that this is not the case.
(Earlier criticisms of this work appeared in \cite{ferreti}.)
Ooguri\cite{private}\ has pointed out that the analysis of \cite{ooguri}\
implies that discrepancies are likely to arise at the level of four graviton
scattering.  The present authors have
isolated the error in \cite{arvind}.
Those authors correctly concluded that
there are no
$1/R^7 r^7$ terms in the matrix model {\it effective action}.
This is in agreement with the calculations of \cite{yoneya},
and also with an argument of Taylor and Van Raamsdonk\cite{wati}.
They went on to argue
that there are no such terms in the matrix model {\it S-matrix}.
This is not the case, as will be explained in \cite{toappear}.
The methods of ref. \cite{arvind} do permit very simple calculations
of terms in the matrix model effective action.
The present authors have verified 
the calculation of \cite{yoneya} using these methods
and have also exhibited
agreement of certain terms in the four (and $n$) graviton
scattering amplitudes.  Other terms
are currently being checked.  The results of this investigation
should appear shortly\cite{toappear}.


\end{document}